\documentclass{article}  

\usepackage{amsmath,amsfonts,amssymb}
\usepackage{graphicx}
\usepackage[colorlinks=true, allcolors=blue]{hyperref}
\usepackage[thmmarks,amsmath]{ntheorem}
\usepackage{mathtools}
\usepackage{authblk}
\usepackage[margin=0.9in]{geometry}
\usepackage{subcaption}

\theoremnumbering{arabic}
\theoremstyle{break}
\theoremsymbol{}
\theoremheaderfont{\normalfont\bfseries}
\theorembodyfont{\upshape}
\newtheorem{theorem}{Theorem}

\newtheorem{proposition}[theorem]{Proposition}
\newtheorem{remark}[theorem]{Remark}

\theoremstyle{nonumberbreak}

\begin{document}

\title{The SWAP Imposter: \\
Bidirectional Quantum Teleportation and its Performance}
\author[1,2]{Aliza U.~Siddiqui}
\affil[1]{\small Division of Computer Science and Engineering, Louisiana State University,
Baton Rouge, Louisiana 70803, USA}
\affil[2]{\small Hearne Institute for Theoretical Physics, Department of Physics and Astronomy, and Center for Computation and Technology, Louisiana State University, Baton Rouge, Louisiana 70803, USA}
\affil[3]{\small School of Electrical and Computer Engineering, Cornell University, Ithaca, New
York 14850, USA}
\author[3,2]{Mark M.~Wilde}

\pagestyle{plain} 

\maketitle

\begin{abstract}
Bidirectional quantum teleportation is a fundamental protocol for exchanging quantum information between two parties. Specifically, the two individuals make use of a shared resource state as well as local operations and classical communication (LOCC) to swap quantum states. In this work, we concisely highlight the  contributions of our companion paper [Siddiqui and Wilde, arXiv:2010.07905]. We develop two different ways of quantifying the error of nonideal bidirectional teleportation by means of the normalized diamond distance and the channel infidelity. We then establish that the values given by both metrics are equal for this task. Additionally, by relaxing the set of operations allowed from LOCC to those that completely preserve the positivity of the partial transpose, we obtain semidefinite programming \textit{lower bounds} on the error of nonideal bidirectional teleportation.
We evaluate these bounds for some key examples---isotropic states and when there is no resource state at all. In both cases, we find an analytical solution. The second example establishes a benchmark for classical versus quantum bidirectional teleportation. Another example that we investigate consists of two Bell states that have been sent through a generalized amplitude damping channel (GADC). For this scenario, we find an analytical expression for the error, as well as a numerical solution that agrees with the former up to numerical precision. 
\end{abstract}

\section{INTRODUCTION}
\label{sec:intro}  

\begin{figure}
\centering
\includegraphics[scale=0.5]{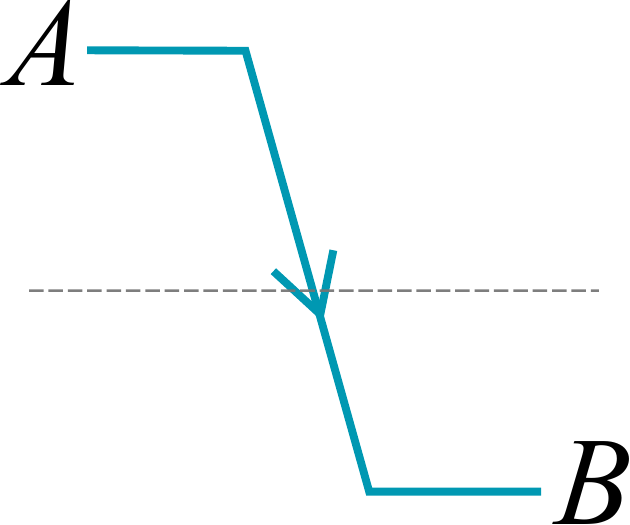}
\caption{Ideally, we would want to send quantum information from one party to another directly via a quantum channel.}
\label{fig: Ideal QT}
\end{figure}

Quantum teleportation is one of the most prominent protocols in quantum information due to its ability to communicate a quantum state between two individuals who share entanglement. In this protocol, there is no need to transmit the physical system. 
While direct transmission of a qubit is possible (shown in Figure \ref{fig: Ideal QT}), its fragile nature is well known. Environmental noise will either corrupt the information encoded in the qubit or prevent it from arriving at its destination altogether. As a result, the quantum teleportation protocol serves as an alternative to physical transmission and utilizes shared entanglement as well as local operations and classical communication (LOCC) to achieve this goal. Teleportation is now commonly used as a fundamental building block in quantum information science, with applications in quantum communication, quantum error correction, quantum networking, etc. 

\begin{figure}
\begin{center}
\includegraphics[scale=0.30]{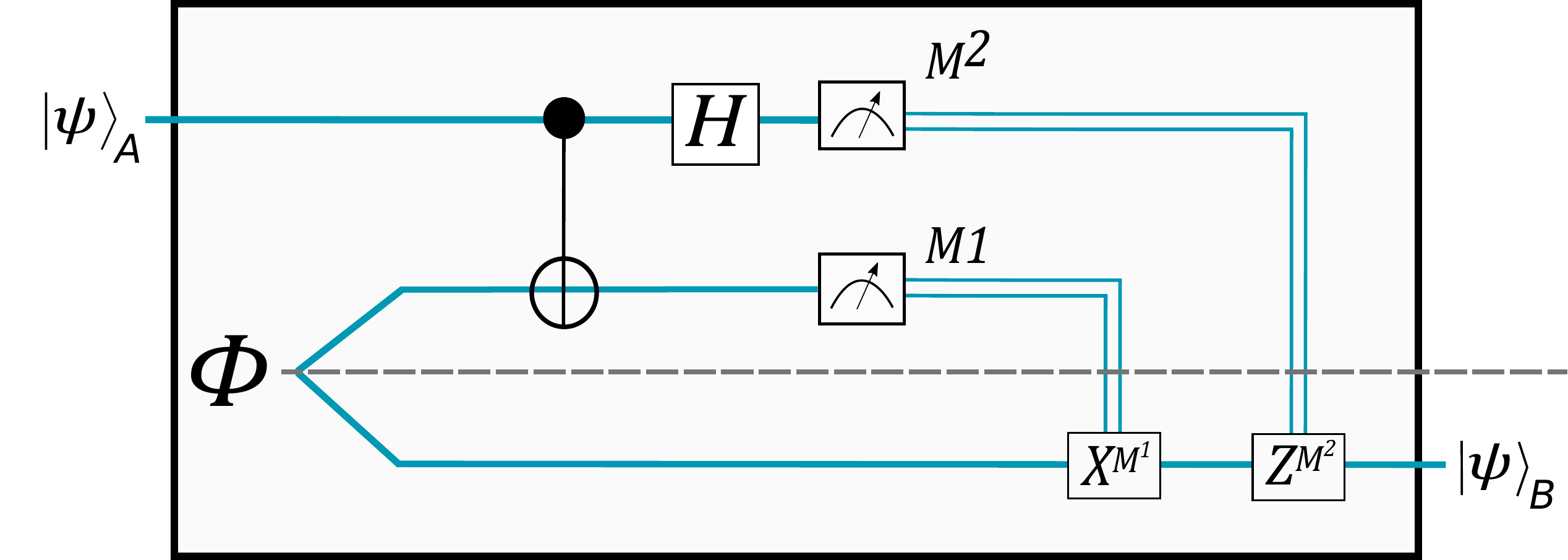}
\end{center}
\caption{Due to the fragile nature of quantum bits, the unidirectional quantum teleportation protocol (shown above) was devised as a method for simulating an ideal unidirectional quantum channel, i.e., to transmit quantum information from one party Alice, to another party Bob.}
\label{fig: Quantum Teleportation Circuit}
\end{figure}

As a reminder, the procedure of standard quantum teleportation is as follows: 
\begin{enumerate}
    \item Two parties, Alice and Bob, are spatially separated and share a maximally entangled state $\Phi_{\hat{A}\hat{B}}$ defined as%
    \begin{equation}
    \label{eq: max-entangled-state}
    \Phi_{\hat{A}\hat{B}}\coloneqq \frac{1}{2}\sum_{i,j=0}^{1}|i\rangle\!\langle j|_{\hat{A}}\otimes|i\rangle\!\langle
    j|_{\hat{B}}.
    \end{equation}
    \item Alice wishes to send her system $A$ to Bob. So, she performs a projective Bell measurement on her systems $A$ and $\hat{A}$.
    \item Alice obtains two classical values from her measurement and transmits them to Bob via a classical communication channel.
    \item Bob, based on the classical results, performs corrective operations on his system of the shared entangled state $\hat{B}$ to recover the original state Alice wished to transfer.
\end{enumerate}
See Figure~\ref{fig: Quantum Teleportation Circuit} for a quantum circuit depiction of the teleportation protocol.

Teleportation has been extended in various ways, and one  way is through bidirectional quantum teleportation. It should be noted that standard quantum teleportation—--also known as \textit{unidirectional teleportation}---realizes a one-way ideal quantum communication channel from one party Alice to another party Bob. The idea of bidirectional teleportation is to provide a two-way quantum communication channel. Instead of only Alice having the ability to transmit quantum information to Bob, individuals can now \textit{exchange} quantum information. In the ideal version of this protocol, our two parties share two pairs of maximally entangled qubits (ebits) and teleport qubits to each other, performing standard quantum teleportation twice in opposite directions. We will refer to any state shared by individuals to perform any variation of teleportation as a \textit{resource state}. The ideal protocol therefore utilizes two ebits as its resource state and is equivalent to a perfect swap channel between two individuals, as shown in Figure~\ref{fig: Ideal BQT Channel}. This extension of teleportation was observed early on in \cite{V94}, and it was subsequently considered in \cite{PhysRevA.63.042303,PhysRevA.65.042316}.
There has recently been a flurry of research on the topic with various proposals for bidirectional teleportation \cite{Fu2014,Hassanpour2016}. There has been even more interest in a variation called bidirectional,
controlled teleportation, using five qubit
\cite{ZZQS13,Shukla2013,Li2013a,Li2013,Chen2014}, six qubit
\cite{Yan2013,Sun2013,Duan2014,Li2016a,ZXL19}, seven qubit
\cite{Duan2014a,Hong2016,Sang2016}, eight qubit
\cite{Zhang2015,SadeghiZadeh2017}, and nine qubit\ \cite{Li2016}\ entangled resource states (see also \cite{Thapliyal2015}). Bidirectional controlled teleportation is a tripartite protocol in which three individuals, typically called Alice, Bob, and Charlie, share an entangled resource state and use LOCC to exchange qubits between Alice and Bob. In other words, Charlie is present to assist Alice and Bob, who wish to swap quantum information. See also \cite{Gou2017} for other variations of bidirectional teleportation.

The applications of bidirectional teleportation align with those of standard teleportation. Specifically, it applies in a basic quantum network setting in which two parties would like to exchange quantum information. 
Although the ideal version of bidirectional teleportation is a trivial extension of the original protocol in which the latter is simply conducted twice (but in opposite directions), the situation becomes less trivial and more relevant to experimental practice when the quantum resource state deviates from the ideal resource of two maximally entangled states. Much of the prior work focuses on precisely this kind of case, when the resource state is different from two maximally entangled states, either by being a different pure state (such as cluster states), a mixed state, or a state with insufficient entanglement to accomplish the task. These kinds of investigations are essential for understanding ways to simulate or mimic the ideal protocol approximately in an experimental setting. 

Despite the many works listed above on the topic of bidirectional teleportation, which try to perform the protocol using various resource states, a systematic method for quantifying its performance has been missing. In other words, there is no procedure to concretely determine how well a certain protocol of bidirectional teleportation performs in comparison to the ideal protocol. How shall we know we have found our perfect imposter?
Any experimental implementation of bidirectional teleportation will necessarily be imperfect and therefore, there is a need for such a metric. Indeed, entangled states generated in experimental settings using methods such as spontaneous parametric down-conversion are only approximations to ideal maximally entangled states \cite{Cout18}. Our aim in \cite{siddiqui2020} and in the present paper is to fill this void.

 While this paper will give basic insight to the work, all proofs as well as additional material are present in our main paper. The present paper instead aims to highlight the essential contributions of our companion paper \cite{siddiqui2020} and give some additional clarification.

\section{PRELIMINARIES}
\label{sec:prelim}

Before proceeding further, we establish some notation and concepts that will be used throughout this paper. In our work, instead of considering only qubits, we generalize all of our scenarios to qudits with dimension $d$.
Specifically, we consider two different dimensions---the dimension of the resource state and the dimension of the unitary swap channel we are trying to mimic. Given two parties Alice and Bob, the dimension of Alice’s qudit that she wishes to send is denoted as $d_{A}$ (similarly $d_{B}$ for Bob). The dimensions of Alice and Bob's qudits are also equivalent to the dimension of the swap channel $d$. The dimension of the shared resource state when Bob’s system is discarded is denoted as $d_{\hat{A}}$ (similarly  $d_{\hat{B}}$ if Alice’s system was discarded). While we consider the case where $d_{A}$ = $d_{B}$ = $d$, we make no assumptions about $d_{\hat{A}}$ and $d_{\hat{B}}$ other than the fact that they are finite-dimensional. In other words, it need not be the case that $d_{\hat{A}}$ = $d_{\hat{B}}$. Additionally, instead of maximally entangled qubits, individuals will now share maximally entangled qudits (e-dits).

We also make use of the following bilateral unitary twirl channel in our paper%
\begin{equation}
\widetilde{\mathcal{T}}_{CD}(X_{CD})\coloneqq \int dU\ (\mathcal{U}_{C}\otimes
\overline{\mathcal{U}}_{D})(X_{CD}), \label{eq:bilat-twirl}%
\end{equation}
where $\mathcal{U}(\cdot) \coloneqq U(\cdot)U^{\dag}$, $\overline{\mathcal{U}}%
(\cdot) \coloneqq \overline{U}(\cdot)U^{T}$ (the overline indicates the complex conjugate), $X_{CD}$ is the bipartite quantum state, and $dU$ denotes the Haar measure (uniform distribution on unitary operators). The bilateral twirl channel is an LOCC channel, in the sense that Alice can pick a unitary at random according to the Haar measure, apply it to her system, report to Bob via a classical channel which one she selected, and Bob can then apply the complex conjugate unitary to his system.

The bilateral twirl is typically utilized to symmetrize quantum states. Specifically, depending on the type of twirl performed, the output state will be invariant under any unitary channel of the form $\mathcal{U} \otimes \overline{\mathcal{U}}$ or the form $\mathcal{U} \otimes \mathcal{U}$. 
For example, given a quantum state $\widetilde{\sigma}_{AB}$ prepared by the \textit{isotropic} bilateral twirl
\begin{equation}
    \int dU\ (\mathcal{U}_{A}\otimes
    \overline{\mathcal{U}}_{B})(\sigma_{AB}) = \widetilde{\sigma}_{AB},
\end{equation}
the following holds for every unitary channel $\mathcal{U}$:
\begin{equation}
    (\mathcal{U_{A}} \otimes \overline{\mathcal{U}}_{B} )   (\widetilde{\sigma}_{AB})    = \widetilde{\sigma}_{AB}.
\end{equation}
Additionally, states prepared by this operation can be described by fewer parameters. Therefore, the twirled state $\widetilde{\sigma}_{AB}$ now has a sparse density matrix and can be characterized by fewer variables. 

\section{IDEAL BIDIRECTIONAL TELEPORTATION}
\label{sec:ideal bqt}

\begin{figure}
\begin{center}
\includegraphics[scale=0.40]{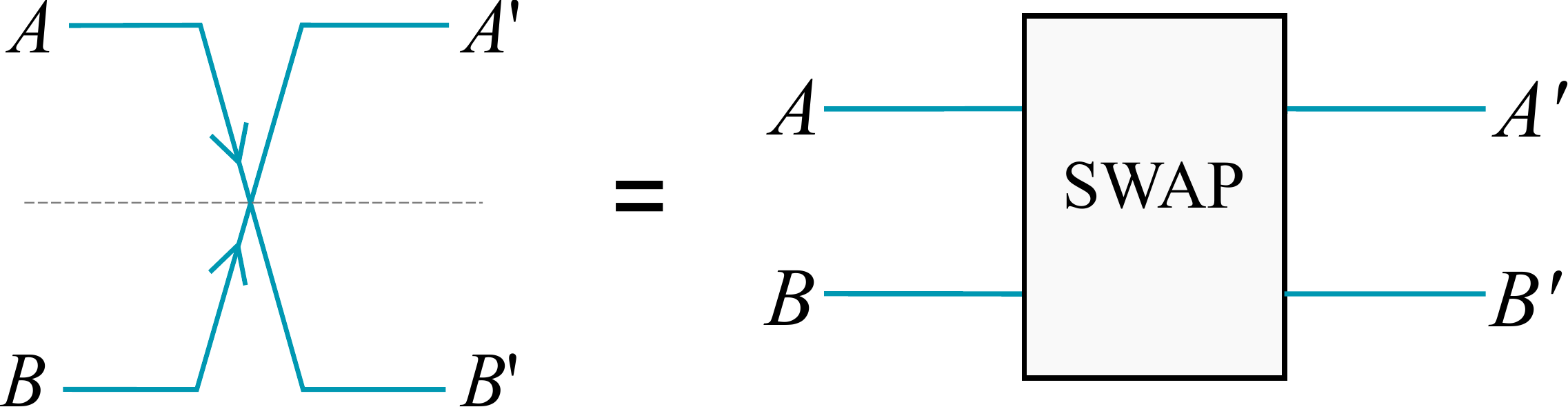}
\end{center}
\caption{Ideal swap channel between two parties, Alice and Bob, realized by ideal bidirectional teleportation. This is a two-party generalization of the ideal unidirectional channel depicted in Figure~\ref{fig: Ideal QT}. Ideal bidirectional quantum teleportation realizes a perfect SWAP channel between two parties.}
\label{fig: Ideal BQT Channel}
\end{figure}

Let us first examine the case of ideal bidirectional teleportation on two qudits in detail. Doing so is helpful in establishing a basic metric for when we consider nonideal bidirectional teleportation later. As stated in the introduction, a trivial way to conduct quantum teleportation bidirectionally between two spatially separated parties, Alice and Bob, is by performing two standard quantum teleportations, once in each direction. This method uses entanglement—--specifically two pairs of e-dits—--as well as local operations and classical communication to behave like a unitary swap channel $\mathcal{S}^{d}_{AB}$ of dimension $d$, shown in Figure \ref{fig: Ideal BQT Channel}.

Since we are examining ideal bidirectional teleportation, each teleportation between the two parties is perfect and equivalent to an identity channel from Alice to Bob and vice versa:
\begin{equation}
\label{eq:swap-channel}
  \mathcal{S}^{d}_{AB} = \operatorname{id}_{A\rightarrow B} \otimes \operatorname{id}_{B\rightarrow A}.
\end{equation}
The proof for \eqref{eq:swap-channel} is outlined in our companion paper \cite{siddiqui2020}.
Even though our choice of notation might suggest that the swap channel is a tensor product of local identity channels, we should note that this is not the case. The swap channel is a global channel that cannot be realized by local actions alone. Our notation $\operatorname{id}_{A\rightarrow B}$ instead indicates that Alice’s input system $A$ is placed at Bob’s output system $B$, and the notation $\operatorname{id}_{B\rightarrow A}$ indicates that Bob’s input system $B$ is placed at Alice’s output system $A$. In other words, Alice's input is perfectly recovered on Bob's end and vice versa. 

\section{QUANTIFYING THE PERFORMANCE OF NONIDEAL BIDIRECTIONAL TELEPORTATION}
\label{sec:unideal bidirectional teleportation}

Now that we have established that ideal bidirectional quantum teleportation is equivalent to the unitary swap channel, it is time to explore nonideal bidirectional quantum teleportation. The goal of nonideal bidirectional teleportation is to become an ``imposter'' of the swap and simulate a $d$-dimensional unitary swap channel as closely as possible using a resource state other than the one required. In other words, the probability of being able to distinguish the swap channel $\mathcal{S}_{AB}$ from the simulation $\widetilde{\mathcal{S}}_{AB}$ should be thus as small as possible. 

The nonideal bidirectional teleportation protocol is as follows: there are two systems $A$ and $B$, that serve as inputs for Alice and Bob respectively. They then act with an LOCC channel $\mathcal{L}_{AB\hat{A}\hat{B}\rightarrow AB}$ on their input systems $A$ and $B$ and their shares of the resource state $\rho_{\hat{A}\hat{B}}$ to produce the output systems $A$ and $B$. The simulation channel is depicted in Figure \ref{fig: swaptoLOCC}.

The expression for the overall channel realized by the simulation is:
\begin{equation}
\widetilde{\mathcal{S}}_{AB}(\omega_{AB})\coloneqq \mathcal{L}_{AB\hat{A}\hat
{B}\rightarrow AB}(\omega_{AB}\otimes\rho_{\hat{A}\hat{B}}),
\label{eq:swap-sim}%
\end{equation}
where $\omega_{AB}$ describes the state of Alice and Bob’s input qudits. 
Note that for the simulation, we allow classical communication between Alice and Bob for free and that $\mathcal{L}_{AB\hat{A}\hat{B}\rightarrow AB}$ can be considered a free channel, as is common in the resource theory of entanglement \cite{Bennet1996, Chitambar2019}.

\subsection{Quantifying Error with Normalized Diamond Distance}
\label{sec: error-diamond-distance}
Let us now discuss how to quantify the simulation
error between the swap channel and the simulation channel. The metric for doing so is the normalized diamond distance
\cite{Kit97}, a standard metric in both quantum computation \cite{Kit97} and quantum information \cite{Watrous2018,W17}. Intuitively, the diamond distance can be thought of as a way to characterize the distinguishability ---or quantify the distance---between two quantum channels. Mathematically speaking, this metric quantifies the maximum absolute deviation between the probabilities of observing the same outcome when each quantum channel is applied to the same input state and the same measurement is made.

\begin{figure}
\begin{center}
\includegraphics[scale=0.55]{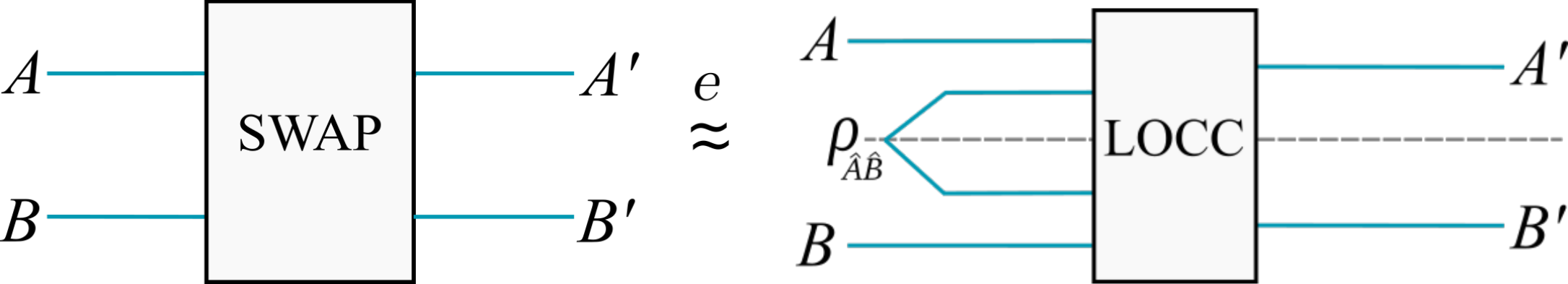}
\end{center}
\caption{The diagram depicts a general framework( shown on the right) for understanding the simulation of bipartite quantum channels, realized by combining an LOCC protocol and  a quantum resource state $\rho_{\hat{A}\hat{B}}$. In experimental implementations, the resource state $\rho_{\hat{A}\hat{B}}$ is imperfect. In our work, we use this framework to simulate the swap channel (left) up to some error $e$.}
\label{fig: swaptoLOCC}
\end{figure}

The formal definition of the normalized diamond distance for two quantum channels $\mathcal{N}$ and $\widetilde{\mathcal{N}}$ is given by
\begin{equation}
\frac{1}{2}\left\Vert \mathcal{N}%
-\widetilde{\mathcal{N}}\right\Vert _{\diamond},
\end{equation}
where the diamond distance 
 $\left\Vert \mathcal{N}-\widetilde{\mathcal{N}%
}\right\Vert _{\diamond}$ is defined as%
\begin{equation}
\left\Vert \mathcal{N}-\widetilde{\mathcal{N}}\right\Vert _{\diamond}%
\coloneqq \sup_{\rho_{RC}}\left\Vert \mathcal{N}_{C\rightarrow D}(\rho_{RC}%
)-\widetilde{\mathcal{N}}_{C\rightarrow D}(\rho_{RC})\right\Vert _{1},
\label{eq:def-diamond-distance}
\end{equation}
and the trace norm of an operator $X$ is given by $\left\Vert X\right\Vert_{1}\coloneqq \operatorname{Tr}[\sqrt{X^{\dag}X}]$. The calculation for the diamond distance in \eqref{eq:def-diamond-distance} can simplify to
\begin{equation}
\left\Vert \mathcal{N}-\widetilde{\mathcal{N}}\right\Vert _{\diamond}%
=\sup_{\psi_{RC}}\left\Vert \mathcal{N}_{C\rightarrow D}(\psi_{RC}%
)-\widetilde{\mathcal{N}}_{C\rightarrow D}(\psi_{RC})\right\Vert _{1},
\label{eq:d-dist-pure-states}%
\end{equation}
where, instead of arbitrary states $\rho_{RC}$, the optimization is with respect to every pure bipartite state $\psi_{RC}$ with system $R$ isomorphic to the channel input system $C$. This simplification is explained in Section~3.5.3 of \cite{KW20book}.  
The normalized diamond distance can be computed through a semidefinite program (definition recalled in Section~IV-A of our companion paper \cite{siddiqui2020}).
Returning to our case of interest, the simulation error of the swap channel, depicted in Figure \ref{fig: swaptoLOCC}, is quantified as follows
\begin{equation}
\label{eq: sim-error-swap}
e_{\operatorname{LOCC}}(\mathcal{S}_{AB}^{d},\rho_{\hat{A}\hat{B}}%
,\mathcal{L}_{AB\hat{A}\hat{B}\rightarrow AB}) \coloneqq \frac{1}{2}\left\Vert
\mathcal{S}^{d}-\widetilde{\mathcal{S}}\right\Vert _{\diamond},
\end{equation}
where $\mathcal{S}^{d}$ is the ideal swap channel and $\widetilde{\mathcal{S}}$ is the simulation channel (see $\mathcal{S}^{d}$ defined in~(\ref{eq:swap-channel}) and $\widetilde{\mathcal{S}}$ defined in~(\ref{eq:swap-sim})).

Since we wish to find the smallest simulation error possible, we minimize \eqref{eq: sim-error-swap} over all LOCC channels. In other words, using the resource state $\rho_{\hat{A}\hat{B}}$, if we substituted an LOCC channel into the bipartite channel simulation framework shown in Figure \ref{fig: swaptoLOCC}, calculated the error, and repeated this procedure for all possible LOCC channels, what is the smallest simulation error we can find? Mathematically, this question is posed as:
\begin{equation}
e_{\operatorname{LOCC}}(\mathcal{S}_{AB}^{d},\rho_{\hat{A}\hat{B}
}) \coloneqq \label{eq:sim-err-swap-DD}\inf_{\mathcal{L}\in\operatorname{LOCC}}e_{\operatorname{LOCC}}(\mathcal{S}_{AB}^{d},\rho_{\hat{A}\hat{B}},\mathcal{L}_{AB\hat{A}\hat{B}\rightarrow AB}).
\end{equation}
This simulation error is difficult to compute as $d$, $d_{\hat{A}}$, and $d_{\hat{B}}$ become larger. 
This computational strain is related to how it is  difficult to optimize over the set of LOCC channels. In Section~\ref{sec:semidefinite prog lower}, we instead determine a \textit{lower bound} on the simulation error which can be computed by means of semidefinite programming and is thus efficiently computable. It will be shown later that for some states $\rho_{\hat{A}\hat{B}}$ of interest, we can determine the error of nonideal bidirectional teleportation exactly. 

\subsection{Quantifying Error with Channel Infidelity}
Another way to quantify error between quantum channels is by utilizing the fidelity measure.
Recall that fidelity of two arbitrary quantum states $\omega$ and $\tau$ is defined as:
\begin{equation}
F(\omega,\tau)\coloneqq \left\Vert \sqrt{\omega}\sqrt{\tau}\right\Vert _{1}^{2}.
\end{equation}

This quantity is equal to one if and only if the states $\omega$ and $\tau$  are the same and is equal to zero if and only if $\omega$ and $\tau$  are orthogonal to each other.
If one of the two states is a pure state, then the definition of fidelity reduces to the following expression:
\begin{equation}
\label{eq: fidelity-with-pure-state}
F(|\psi\rangle\!\langle\psi|,\tau)=\langle\psi|\tau|\psi\rangle.
\end{equation}
Intuitively, fidelity measures the amount of overlap two quantum states have with each other. 
In the case of \eqref{eq: fidelity-with-pure-state}, it has the operational meaning that $F(|\psi\rangle\!\langle
\psi|,\tau)$ is the probability with which the state $\tau$ passes a test for
being the state $|\psi\rangle\!\langle\psi|$. The test in this case is given
by the binary measurement $\{|\psi\rangle\!\langle\psi|,I-|\psi\rangle
\!\langle\psi|\}$, and the first outcome corresponds to the decision
\textquotedblleft pass\textquotedblright.\ So the probability of passing is
equal to $F(|\psi\rangle\!\langle\psi|,\tau)$.

\begin{figure}[t]
\begin{center}
\includegraphics[scale=0.40]{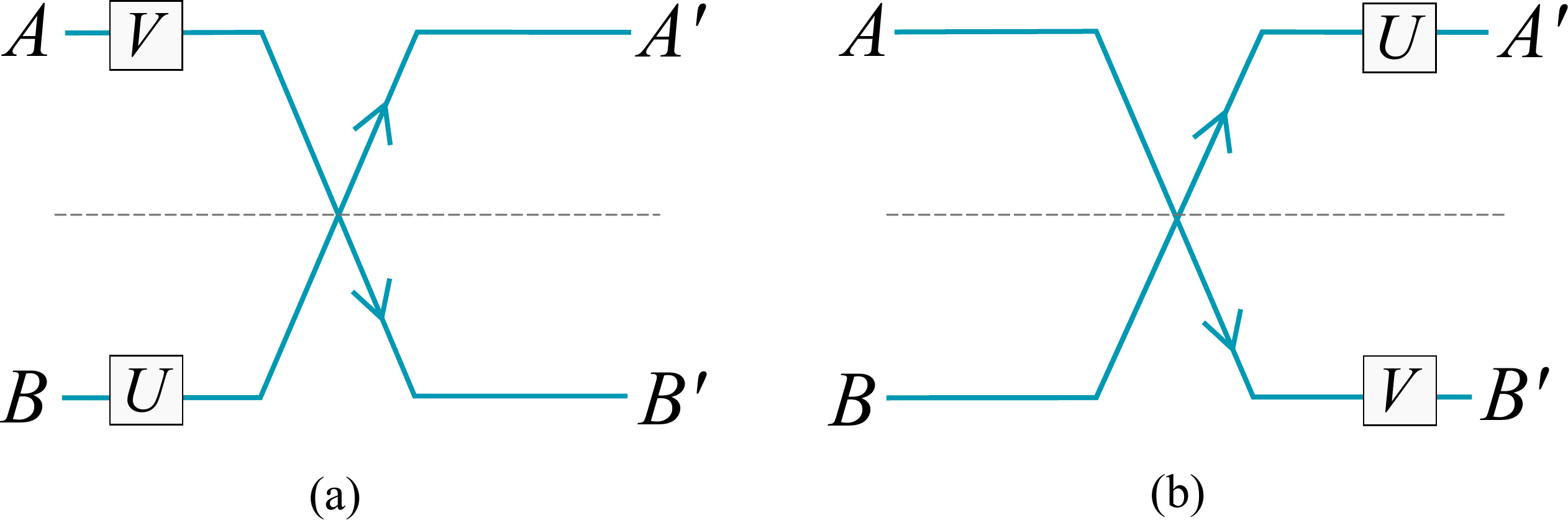}
\end{center}
\caption{The symmetries of the SWAP channel are depicted
in this figure. By exploiting this unique property, (a) and (b) above are equivalent and the optimization problem for quantifying the performance of unideal bidirectional teleportation can be greatly simplified. (a) Alice and Bob perform unitary operations V and U, respectively, and then a SWAP operation to exchange information. (b) Alice and Bob first perform a SWAP operation and then unitary operations U
and V , respectively, on their individual qubits.}
\label{fig: symmetry-of-swap}
\end{figure}

We can now extend the fidelity measure of quantum states to measure the similarity between two quantum channels $\mathcal{N}_{C\rightarrow D}$ and  $\widetilde{\mathcal{N}}_{C\rightarrow D}$ as follows:
\begin{equation}
F(\mathcal{N},\widetilde{\mathcal{N}}) \coloneqq  \inf_{\rho_{RC}}F(\mathcal{N}%
_{C\rightarrow D}(\rho_{RC}),\widetilde{\mathcal{N}}_{C\rightarrow D}%
(\rho_{RC})),
\end{equation}
This expression can be viewed as the fidelity counterpart of the diamond distance in
\eqref{eq:def-diamond-distance}. Just like \eqref{eq:d-dist-pure-states}, the
following simplification holds%
\begin{equation}
F(\mathcal{N},\widetilde{\mathcal{N}})=\inf_{\psi_{RC}}F(\mathcal{N}%
_{C\rightarrow D}(\psi_{RC}),\widetilde{\mathcal{N}}_{C\rightarrow D}%
(\psi_{RC})),
\end{equation}
where the optimization is with respect to all pure bipartite states $\psi
_{RC}$ with system $R$ isomorphic to the channel input system $C$. In our work, we make use of channel \textit{infidelity} which will quantify the error between the two quantum channels $\mathcal{N}_{C\rightarrow D}$ and $\widetilde{\mathcal{N}}_{C\rightarrow D}$. The channel infidelity is defined as
\begin{equation}
1 - F(\mathcal{N},\widetilde{\mathcal{N}}).
\end{equation}
(In Section IVB of our main paper \cite{siddiqui2020}, we recall a semi-definite program to calculate the root fidelity of quantum channels).  
Using the fidelity of channels, we can define an alternate notion of
simulation error as%
\begin{equation}
e_{\operatorname{LOCC}}^{F}(\mathcal{S}_{AB}^{d},\rho_{\hat{A}\hat{B}%
},\mathcal{L}_{AB\hat{A}\hat{B}\rightarrow AB})\coloneqq 1-F(\mathcal{S}%
^{d},\widetilde{\mathcal{S}}),
\end{equation}
where $d$ is the dimension of the swap channel $\mathcal{S}^{d}$. Minimizing this error with respect to all LOCC channels,
we arrive at the following:%
\begin{equation}
e_{\operatorname{LOCC}}^{F}(\mathcal{S}_{AB}^{d},\rho_{\hat{A}\hat{B}%
})\coloneqq \label{eq:sim-err-swap-ch-infid}
\inf_{\mathcal{L}\in\operatorname{LOCC}}e_{\operatorname{LOCC}}^{F}%
(\mathcal{S}_{AB}^{d},\rho_{\hat{A}\hat{B}},\mathcal{L}_{AB\hat{A}\hat
{B}\rightarrow AB}).
\end{equation}
For the same reasons given previously in Section~\ref{sec: error-diamond-distance}, this quantity is difficult to compute,
and so we seek alternative ways to estimate it.

\begin{remark}
Even though we have defined two
different notions of LOCC\ simulation error of bidirectional teleportation based on the normalized diamond distance and channel infidelity, the error values they give are equivalent for the simulation of the swap channel. This result follows as a consequence of the swap channel $\mathcal{S}_{AB}^{d}$ in \eqref{eq:swap-channel} having the following
symmetry (shown pictorially in Figure \ref{fig: symmetry-of-swap})%
\begin{equation}
\left( \mathcal{V}_{A}\otimes\mathcal{U}_{B}\right)  \mathcal{S}_{AB}^{d} = \mathcal{S}_{AB}^{d}\left(  \mathcal{U}_{A}\otimes\mathcal{V}_{B}\right) ,
\label{eq:swap-ch-symmetry-for-LOCC}%
\end{equation}
holding for all unitary channels $\mathcal{U}_{A}$ and $\mathcal{V}_{B}$. The proof of the claim can be found in Appendix~A of our main paper \cite{siddiqui2020}.
\end{remark}

\section{SEMI-DEFINITE PROGRAMMING LOWER BOUNDS}
\label{sec:semidefinite prog lower}

\begin{figure}
    \centering
    \begin{subfigure}{0.25\textwidth}
    \centering
    \includegraphics[scale=0.28]{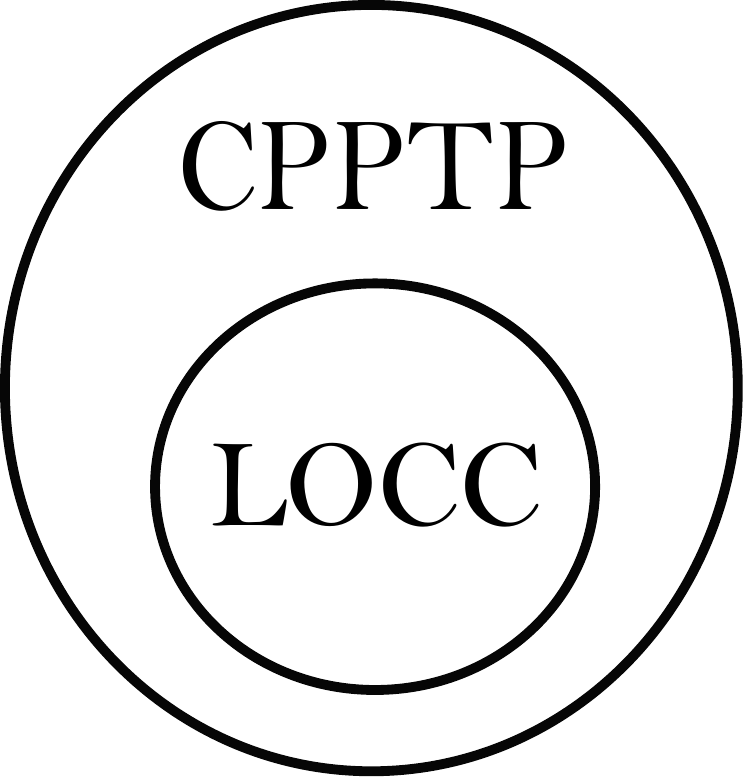}
    \caption{}
    \label{fig:cpptpLOCC}
    \end{subfigure}
    \qquad
    \begin{subfigure}{0.36\textwidth}
    \includegraphics[scale=0.4]{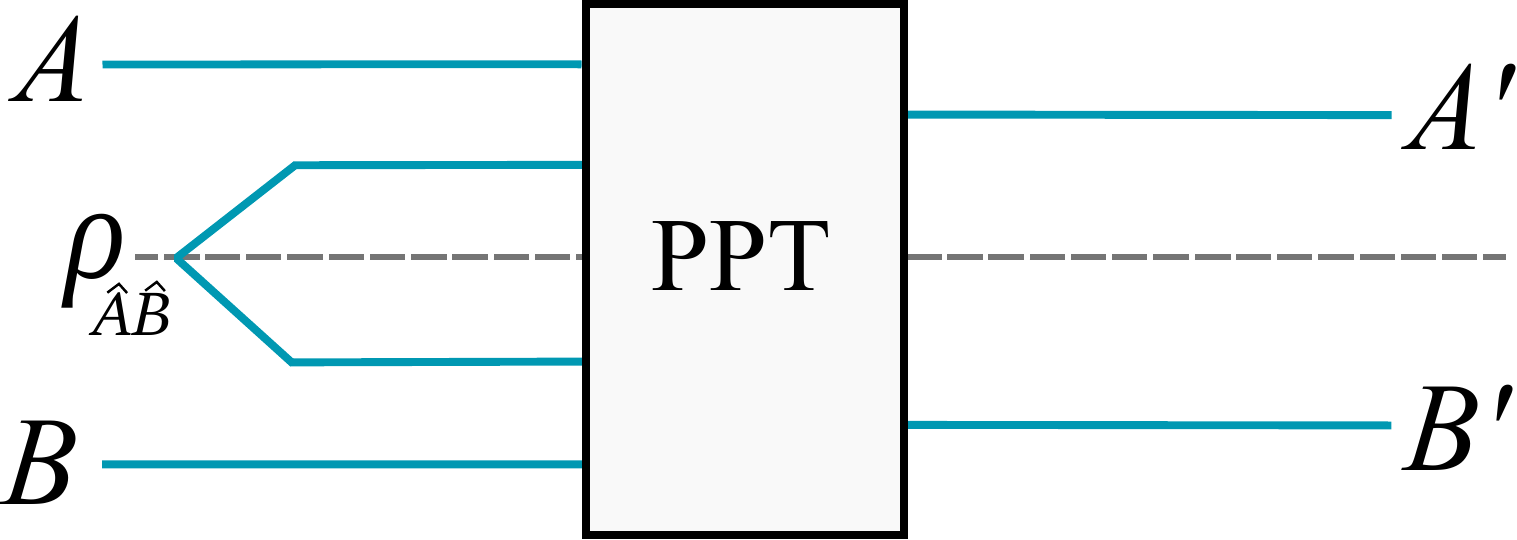}
    \caption{}
    \label{fig:unidealPPT}
    \end{subfigure}
    \caption{Optimizing over all LOCC channels is known to be computationally intensive. (a) We utilize the fact that LOCC channels are a subset of channels that completely preserve the positivity of the partial transpose (C-PPT-P).  (b) Instead of optimizing over all LOCC channels and the ensuing protocols depicted in Figure~\ref{fig: swaptoLOCC}, we relax the optimization to the larger set of C-PPT-P channels. Conducting the optimization problem over this larger set can be solved in time polynomial in the dimension of the resource state and the swap channel to be simulated.}%
    \label{fig:relaxLOCC}%
\end{figure}


As discussed before, it is challenging to compute the simulation errors mentioned in \eqref{eq:sim-err-swap-DD} and \eqref{eq:sim-err-swap-ch-infid} because it is difficult to optimize over the set of LOCC channels. 
Consequently, we follow the approach of \cite{Rai99, Rai01} and enlarge our optimization set from the set of all LOCC channels to the larger set of completely positive-partial-transpose-preserving channels instead (denoted as C-PPT-P or PPT for short).
We can then optimize with respect to this superset and alternatively, obtain a lower bound on the errors in \eqref{eq:sim-err-swap-DD} and \eqref{eq:sim-err-swap-ch-infid}. 
First, recall that a bipartite channel $\mathcal{N}_{AB\rightarrow A^{\prime
}B^{\prime}}$ is defined to be C-PPT-P \cite{Rai99, Rai01} if the map
\begin{equation}
T_{B^{\prime}}\circ\mathcal{N}_{AB\rightarrow A^{\prime
}B^{\prime}}\circ
T_{B}, \label{eq:C-PPT-P-condition}%
\end{equation}
is completely positive, where $T_{B}$ denotes the transpose map, defined by%
\begin{equation}
T_{B}(\omega_{B})=\sum_{i,j}|i\rangle\!\langle j|_{B}\omega_{B}|i\rangle
\!\langle j|_{B}, \label{eq:transpose-map}%
\end{equation}
and with $T_{B^{\prime}}$ defined similarly on the system $B^{\prime}$.
According to \cite{Rai99}, the set of all LOCC channels is a subset of the set of channels that completely preserve the positive partial transpose but not every C-PPT-P channel is an LOCC channel.
Therefore,    
\begin{equation}
\text{LOCC\ }\subset\ \text{C-PPT-P}, \label{eq:LOCC-in-CPPTP}%
\end{equation}
as depicted in Figure \ref{fig:cpptpLOCC}. 
Now that we have defined what C-PPT-P channels are, let us redefine the simulation error over these channels based on diamond distance as
\begin{equation}
e_{\operatorname{PPT}}(\mathcal{S}_{AB}%
^{d}%
,\rho_{\hat{A}\hat{B}})\coloneqq \label{eq:c-ppt-p-error}\\
\frac{1}{2}\inf_{\mathcal{P}\in\text{C-PPT-P}}\left\Vert \mathcal{S}%
_{AB\rightarrow A^{\prime}B^{\prime}}-\widetilde{\mathcal{S}}_{AB\rightarrow
A^{\prime}B^{\prime}}\right\Vert _{\diamond},
\end{equation}
where the optimization is with respect to C-PPT-P channels $\mathcal{P}%
_{AB\hat{A}\hat{B}\rightarrow A^{\prime}B^{\prime}}$, and the simulation channel is now defined as%
\begin{equation}
\widetilde{\mathcal{S}}_{AB\rightarrow A^{\prime}B^{\prime}}(\omega
_{AB})\coloneqq \mathcal{P}_{AB\hat{A}\hat{B}\rightarrow A^{\prime}B^{\prime}}%
(\omega_{AB}\otimes\rho_{\hat{A}\hat{B}}).
\end{equation}
Furthermore, as a result of the containment in \eqref{eq:LOCC-in-CPPTP}, the error in simulating the swap channel when optimizing over PPT channels serves as a lower bound on the error when optimizing over LOCC:%
\begin{equation}
e_{\operatorname{PPT}}(\mathcal{S}_{AB\rightarrow A^{\prime}B^{\prime}}%
,\rho_{AB})\leq e_{\operatorname{LOCC}}(\mathcal{S}_{AB\rightarrow A^{\prime
}B^{\prime}},\rho_{AB}). \label{eq:ppt-err-locc-err}%
\end{equation}
 Let us now discuss how the error defined in \eqref{eq:c-ppt-p-error} can be computed by means of a semi-definite program. To do so, we apply semi-definite constraints for optimization over C-PPT-P channels and utilize the symmetry property of the SWAP channel given in \eqref{eq:swap-ch-symmetry-for-LOCC}, as well as the fact that it commutes with itself, to produce the following semi-definite program.

\begin{proposition}

\label{prop:swap-sdp-simplify}
The semi-definite program for the error in simulating the unitary SWAP channel $\mathcal{S}_{AB}^{d}$ in \eqref{eq:swap-channel} by using a resource state $\rho_{\hat{A}\hat{B}}$ and an arbitrary  C-PPT-P channel
is as follows:
\begin{equation}
e_{\operatorname{PPT}}(\mathcal{S}_{AB}^{d},\rho_{\hat{A}\hat{B}}%
)=1-\sup_{\substack{K_{\hat{A}\hat{B}},L_{\hat{A}\hat{B}},\\N_{\hat{A}\hat{B}%
}\geq0}}\operatorname{Tr}[\rho_{\hat{A}\hat{B}}K_{\hat{A}\hat{B}}],
\label{eq:main-SDP-paper}%
\end{equation}
subject to
\begin{align}
T_{\hat{B}}\!\left(  K_{\hat{A}\hat{B}}+\frac{L_{\hat{A}\hat{B}}}{d+1}%
+\frac{N_{\hat{A}\hat{B}}}{\left(  d+1\right)  ^{2}}\right)   &
\geq0,\nonumber\\
\frac{1}{d^{2}-1}T_{\hat{B}}\!\left(  L_{\hat{A}\hat{B}}+N_{\hat{A}\hat{B}%
}\right)   &  \geq T_{\hat{B}}\!\left(  K_{\hat{A}\hat{B}}\right)
,\nonumber\\
T_{\hat{B}}\!\left(  K_{\hat{A}\hat{B}}+\frac{N_{\hat{A}\hat{B}}}{\left(
d-1\right)  ^{2} }\right)   &  \geq\frac{1}{d-1}T_{\hat{B}}\!\left(
L_{\hat{A}\hat{B}}\right)  ,\nonumber\\
K_{\hat{A}\hat{B}}+L_{\hat{A}\hat{B}}+N_{\hat{A}\hat{B}}  &  =I_{\hat{A}%
\hat{B}}.
\end{align}
where $K_{\hat{A}\hat{B}}$, $L_{\hat{A}\hat{B}}$,  and $N_{\hat{A}\hat{B}}$ are positive semi-definite Hermitian matrices and elements of a POVM and $d$ is the dimension of the SWAP channel to be simulated.  
\end{proposition}

The proof for Proposition~\ref{prop:swap-sdp-simplify}
is given in our companion paper \cite{siddiqui2020}. 

\begin{remark}
\label{rem:channel-form}
Given the proof of Proposition~\ref{prop:swap-sdp-simplify},%
\ an optimal C-PPT-P\ channel for simulating the unitary
swap channel has the following structure:\begin{multline}
\mathcal{P}_{AB\hat{A}\hat{B}\rightarrow AB}(\omega_{AB}\otimes\rho_{\hat
{A}\hat{B}})=\mathcal{S}_{AB}^{d}(\omega_{AB})\operatorname{Tr}[K_{\hat{A}%
\hat{B}}\rho_{\hat{A}\hat{B}}]\\
+\frac{1}{2}\left(
\operatorname{id}_{A\rightarrow B}\otimes\mathcal{D}_{B\rightarrow A}
+\mathcal{D}_{A\rightarrow B}\otimes\operatorname{id}_{B\rightarrow A}%
\right)  (\omega_{AB})\operatorname{Tr}[L_{\hat{A}\hat{B}}\rho_{\hat{A}\hat
{B}}]\\
+\left(  \mathcal{D}_{A\rightarrow B}\otimes\mathcal{D}_{B\rightarrow
A}\right)  (\omega_{AB})\operatorname{Tr}[N_{\hat{A}\hat{B}}\rho_{\hat{A}%
\hat{B}}],
\end{multline}
where $\mathcal{D}$ denotes the following generalized Pauli channel:%
\begin{equation}
\mathcal{D}(\sigma)\coloneqq\frac{1}{d^{2}-1}\sum_{\left(  x,z\right)
\neq\left(  0,0\right)  }W^{z,x}\sigma(W^{z,x})^{\dag}%
,\label{eq:gen-Pauli-channel}
\end{equation}
and $W^{z,x}$ is a generalized Pauli operator (Heisenberg--Weyl).
Thus, the interpretation of the simulating channel is that it measures the
resource state $\rho_{\hat{A}\hat{B}}$ according to the POVM\ $\{K_{\hat
{A}\hat{B}},L_{\hat{A}\hat{B}},N_{\hat{A}\hat{B}}\}$, which is subject to the
inequality constraints in Proposition~\ref{prop:swap-sdp-simplify}. After
that, it takes the following action:

\begin{enumerate}
\item If the first outcome $K_{\hat{A}\hat{B}}$ occurs, then apply the ideal
swap channel to the input state $\omega_{AB}$.

\item If the second outcome $L_{\hat{A}\hat{B}}$ occurs, then with probability
1/2, apply the identity channel $\operatorname{id}_{A\rightarrow B}$ to
transfer Alice's input system $A$ to Bob, but then garble Bob's input system
$B$ by applying the channel $\mathcal{D}$ and transfer the resulting system to
Alice; with probability 1/2, apply the identity channel $\operatorname{id}%
_{B\rightarrow A}$ to transfer Bob's input system $B$ to Alice, but then
garble Alice's input system $A$ by applying the channel $\mathcal{D}$ and
transfer the resulting system to Bob.

\item If the third outcome $N_{\hat{A}\hat{B}}$ occurs, then apply the
garbling channel $\mathcal{D}$ to both Alice and Bob's systems individually
and exchange them.
\end{enumerate}

\noindent The fact that the measurement operators obey the inequality
constraints in Proposition~\ref{prop:swap-sdp-simplify} implies that the
quantum channel $\mathcal{P}_{AB\hat{A}\hat{B}\rightarrow AB}$ is C-PPT-P.
 We have now found our PPT imposter channel for the swap! 
\end{remark}

It should be briefly noted that in our companion paper \cite{siddiqui2020}, we derive a semi-definite program to calculate the error for simulating the swap channel in terms of channel infidelity. It turns out that the optimal value of this semi-definite program simplifies to the expression from Proposition~\ref{prop:swap-sdp-simplify}. Consequently, there is no need to consider different notions of simulation error when considering the simulation of the unitary swap channel using C-PPT-P channels. That being said, it is not necessarily true that these two distance metrics lead to the same simulation error or even the same semi-definite program when the goal is to simulate a general bipartite channel other than the swap channel. 

\section{EXAMPLES}

In this section, we consider several examples of resource states that can be used for bidirectional teleportation, and using the SDP established in Proposition~\ref{prop:swap-sdp-simplify}, we evaluate the performance of the protocol when it employs these states. For several cases of interest, we establish an exact evaluation not only for the error when using a PPT simulation but also when using an LOCC simulation.

\subsection{No Resource State: Benchmark for Classical vs Quantum Bidirectional Teleportation}

\begin{figure}
\begin{center}
\includegraphics[width=10cm]{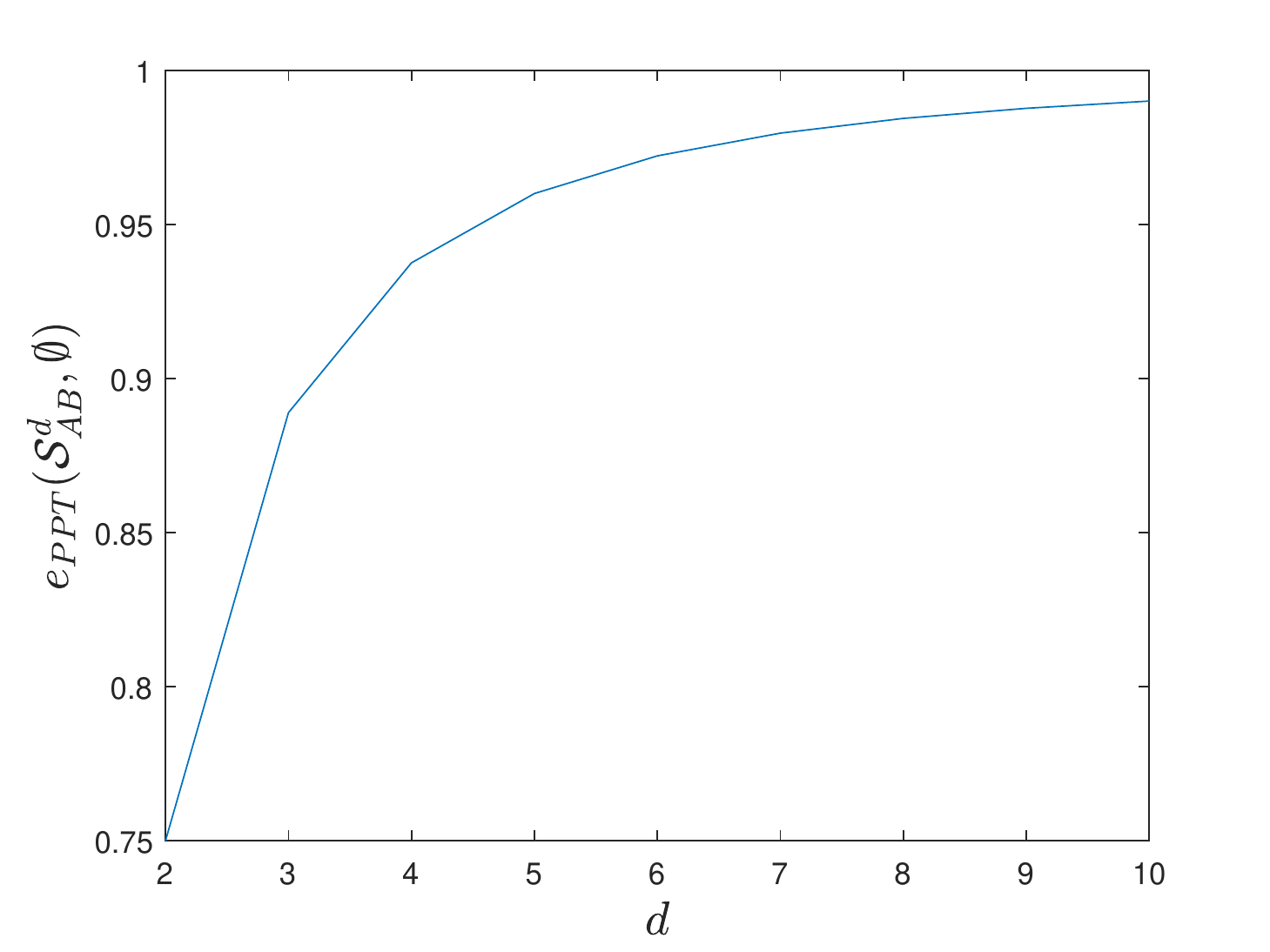}
\end{center}
\caption{Plot of the simulation error of bidirectional teleportation when
using no quantum resource state or,  equivalently, since we allow LOCC for free, a separable state. The quantity $d$ is the dimension of the SWAP channel.}%
\label{fig:noResourceError}%
\end{figure}

One key example of a resource state is when there is in fact no resource state at all. In other words, what if the two parties Alice and Bob did not share any quantum resource state or, in another case, a separable state? How well could they perform bidirectional teleportation?
In this case, both parties can only employ a PPT or LOCC simulation of bidirectional teleportation and can prepare a separable state for free. 
\begin{proposition}
\label{prop:no-res-sim-err}If there is no resource state, then the error in simulating the unitary SWAP channel $\mathcal{S}_{AB}^{d}$ in \eqref{eq:swap-channel} is equal to $1-1/d^{2}$, i.e.,
\begin{equation}
e_{\operatorname{PPT}}(\mathcal{S}_{AB}^{d},\emptyset)=e_{\operatorname{LOCC}%
}(\mathcal{S}_{AB}^{d},\emptyset)=1-\frac{1}{d^{2}}, \label{eq:err-no-res-thm}%
\end{equation}
where the notation $\emptyset$ indicates the absence of a resource state.
\end{proposition}
 Figure~\ref{fig:noResourceError} plots the expression in \eqref{eq:err-no-res-thm} for the simulation error. The importance of this result is that it establishes a worst-case scenario for performing bidirectional teleportation. It creates a dividing line between a classical and a quantum implementation of this protocol, which can be used by experimentalists to assess the performance of an implementation of bidirectional teleportation. \textit{Essentially, all bidirectional teleportation protocols must have a performance with an error less than the bound above to be considered useful}.
 The bound $1-1/d^{2}$ acts as both an upper and lower bound on $e_{\operatorname{PPT}}(\mathcal{S}_{AB}^{d},\emptyset)$ as well as $e_{\operatorname{LOCC}%
}(\mathcal{S}_{AB}^{d},\emptyset)$. The proof for this can be found in our companion paper \cite{siddiqui2020}. 
Note that, while this error holds true for the case where the resource state is a separable or non-entangled state, the error can become smaller if the parties were to share an entangled state (as one would expect).

\subsection{Isotropic States}

Another class of bipartite states we are interested in studying are isotropic states. A fascinating fact is that any arbitrary
state of systems $\hat{A}\hat{B}$ can be twirled to an isotropic state using the channel in~(\ref{eq:bilat-twirl}), making the evaluation of this class particularly important. These states are characterized by two parameters, the fidelity to the maximally entangled state $F\in\left[  0,1\right]  $ as well as the dimension $d_{\hat{A}}\in\left\{  2,3,4,\ldots\right\}$. These states are defined as follows:
\begin{equation}
\rho_{\hat{A}\hat{B}}^{(F,d_{\hat{A}})}\coloneqq F\Phi_{\hat{A}\hat{B}}+\left(
1-F\right)  \frac{I_{\hat{A}\hat{B}}-\Phi_{\hat{A}\hat{B}}}{d_{\hat{A}}^{2}%
-1}. 
\label{eq:isotropic-def}%
\end{equation}
(Recall the definition for a maximally entangled state $\Phi_{\hat{A}\hat{B}}$ given in~(\ref{eq: max-entangled-state})). 

\begin{figure}[ptb]
\begin{center}
\includegraphics[
width=10cm
]{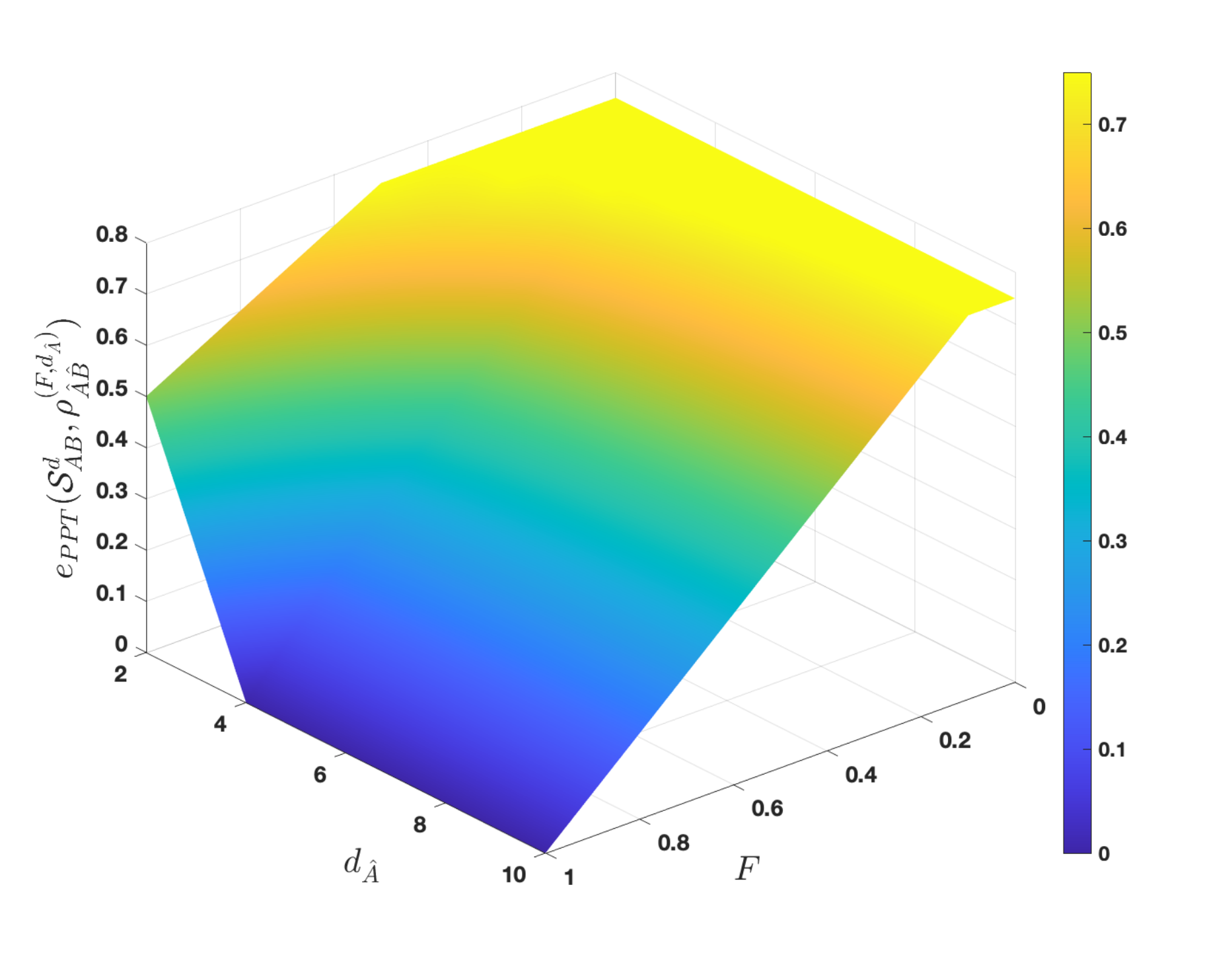}
\end{center}
\caption{Plot of the simulation error of bidirectional teleportation when
using the isotropic resource state defined in \eqref{eq:isotropic-def}, where $F$ is the fidelity parameter and $d_{\hat{A}}$ is the dimension of Alice's system of the resource state.}%
\label{fig:isotropicError}%
\end{figure}

The following proposition establishes a simple expression for the simulation error when using an isotropic state for bidirectional teleportation. It is given exclusively in terms of the dimension $d$ of the swap channel that is
being simulated and the two parameters $F$ and  $d_{\hat{A}}$ that characterize the isotropic resource state. A proof is available in Appendix~E of our companion paper \cite{siddiqui2020}. The proof exploits the symmetries
of an isotropic state (similar to the symmetries of the swap channel shown in Figure~\ref{fig: symmetry-of-swap})  to reduce the semi-definite program in Proposition~\ref{prop:swap-sdp-simplify} to a linear program, which we then solve analytically.
\begin{proposition}
\label{prop:isotropic-sim-perf}
The simulation error for the unitary swap
channel over all PPT channels when using an isotropic resource state $\rho_{\hat{A}\hat{B}%
}^{(F,d_{\hat{A}})}$ is%
\begin{equation}
\label{eq:isotropic-error}
e_{\operatorname{PPT}}(\mathcal{S}_{AB}^{d},\rho_{\hat{A}\hat{B}}%
^{(F,d_{\hat{A}})})=\\
\left\{
\begin{array}
[c]{cc}%
1-\frac{1}{d^{2}} & \text{if }F\leq\frac{1}{d_{\hat{A}}}\\
1-\frac{Fd_{\hat{A}}}{d^{2}} & \text{if }F>\frac{1}{d_{\hat{A}}}\text{ and
}d_{\hat{A}}\leq d^{2}\\
\frac{\left(  1-\frac{1}{d^{2}}\right)  \left(  1-F\right)  }{1-\frac
{1}{d_{\hat{A}}}} & \text{if }F>\frac{1}{d_{\hat{A}}}\text{ and }d_{\hat{A}%
}>d^{2}%
\end{array}
\right.  .
\end{equation}
We also have that%
\begin{equation}
e_{\operatorname{PPT}}(\mathcal{S}_{AB}^{d},\rho_{\hat{A}\hat{B}}%
^{(F,d_{\hat{A}})})=e_{\operatorname{LOCC}}(\mathcal{S}_{AB}^{d},\rho_{\hat
{A}\hat{B}}^{(F,d_{\hat{A}})}) \label{eq:ppt-err-equals-locc-err-iso}%
\end{equation}
if $F\leq\frac{1}{d_{\hat{A}}}$ or if $F>\frac{1}{d_{\hat{A}}}$ and
$d_{\hat{A}}\leq d^{2}$.
\end{proposition}
It is an open question to determine if the equality in
\eqref{eq:ppt-err-equals-locc-err-iso} holds when $F>\frac{1}{d_{\hat{A}}}$ and $d_{\hat{A}}>d^{2}$. 
Figure~\ref{fig:isotropicError} plots the expression given in \eqref{eq:isotropic-error} for the simulation error.

It should be noted that an isotropic state can in fact realize ideal bidirectional quantum teleportation as described in Section~\ref{sec:ideal bqt} when its fidelity $F$ is one. As shown in Figure~\ref{fig:isotropicError}, when fidelity is one and the dimension of the isotropic state  $d_{\hat{A}}$ reaches four and beyond, it encompasses multiple e-dits which Alice and Bob can locally separate out to perform bidirectional teleportation. In the case of $d_{\hat{A}}$ = 4, for example, the two parties can separate out $\Phi_{\hat{A}\hat{B}%
}^{4}$ to two e-dits $\Phi_{A_{1}B_{1}}^{2}\otimes\Phi_{A_{2}B_{2}}^{2}$, and they are then able to perform the ideal bidirectional protocol. As a result, for the case of $F$ = 1 and $d_{\hat{A}}\in\left\{ 4, 5, 6,\ldots\right\}$, the error in simulating the swap channel with the isotropic state $e_{\operatorname{PPT}}(\mathcal{S}_{AB}^{d},\rho_{\hat{A}\hat{B}}^{(F,d_{\hat{A}})})$ reduces to zero. A perfect imposter needs the perfect disguise, which isotropic states indeed prove to be in the limit $F \to 1$.

\subsection{Resource State Resulting from Generalized Amplitude Damping Channel}
In this section, we consider a numerical example in which we can apply the
semi-definite program from Proposition~\ref{prop:swap-sdp-simplify}. This
example involves a resource state resulting from two Bell states affected by
noise from a generalized amplitude damping channel (GADC), a common scenario in most experimental settings. The GADC can be understood as a qubit thermal channel, in which the input qubit interacts with a thermal qubit environment according to a beamsplitter-like interaction, after which the environment qubit is discarded \cite{KSW19}. Intuitively, it models energy relaxation from the excited state to the ground state. In more detail,
recall that the GADC has the following form (see, e.g., \cite{KSW19}):%
\begin{equation}
\mathcal{A}_{\gamma,N}(\rho)\coloneqq \sum_{i=1}^{4}A_{i}\rho A_{i}^{\dag},
\end{equation}
where $\gamma\in\left[  0,1\right]  $ is the damping parameter,
$N\in\left[  0,1\right]  $ is the noise parameter, and the Kraus operators are defined as %
\begin{align}
A_{1}  &  \coloneqq \sqrt{1-N}\left(  |0\rangle\!\langle0|+\sqrt{1-\gamma}%
|1\rangle\!\langle1|\right)  ,\\
A_{2}  &  \coloneqq \sqrt{\gamma\left(  1-N\right)  }|0\rangle\!\langle1|,\\
A_{3}  &  \coloneqq \sqrt{N}\left(  \sqrt{1-\gamma}|0\rangle\!\langle0|+|1\rangle
\!\langle1|\right)  ,\\
A_{4}  &  \coloneqq \sqrt{\gamma N}|1\rangle\!\langle0|.
\end{align}
The resource state we consider is then:%
\begin{equation}
\mathcal{A}_{\gamma,N}^{\otimes4}(\Phi^{\otimes2}). \label{eq:GADC-res-state}%
\end{equation}
The resource state in \eqref{eq:GADC-res-state} is equivalent to two ebits,
consisting of four qubits in total, each of which is acted upon by a GADC with
the same parameters $\gamma$ and $N$. When $\gamma$ and $N$ are both equal to
zero, the resource state is equivalent to two ebits and perfect bidirectional
teleportation is possible. As the noise parameters increase, the bidirectional
teleportation is imperfect and occurs with some error.

By evaluating the semi-definite program in
Proposition~\ref{prop:swap-sdp-simplify}\ for this resource state, we obtain a
lower bound on the simulation error of bidirectional teleportation. We 
obtain an upper bound  by demonstrating a protocol that
uses this resource state. If Alice and Bob perform a bilateral twirl on their
state---specifically, the channel in \eqref{eq:bilat-twirl}---where $U$ is a
unitary that acts on two qubits, then the resulting state is an isotropic
state of the form:%
\begin{equation}
F(\gamma,N)\Phi^{\otimes2}+\left(  1-F(\gamma,N)\right)  \frac{I^{\otimes
4}-\Phi^{\otimes2}}{15},
\end{equation}
where%
\begin{align}
F(\gamma,N) &  \coloneqq \operatorname{Tr}[\Phi^{\otimes2}\mathcal{A}_{\gamma
,N}^{\otimes4}(\Phi^{\otimes2})]\\
&  =\left[  1+\frac{\gamma}{2}\left(  \gamma-2\left[  1+\gamma N\left(
1-N\right)  \right]  \right)  \right]  ^{2}.
\end{align}
By applying Proposition~\ref{prop:isotropic-sim-perf}\ and noting that
$d_{\hat{A}}=d^{2}=4$ for this example, we find that the simulation error,
when using this protocol is given by%
\begin{equation}
1-\max\left\{  F(\gamma,N),\frac{1}{16}\right\}  .\label{eq:upper-bnd-GADC}%
\end{equation}
Up to numerical precision, we find that the upper bound in
\eqref{eq:upper-bnd-GADC} and the SDP\ lower bound from
Proposition~\ref{prop:swap-sdp-simplify}\ match, so that
\eqref{eq:upper-bnd-GADC} should in fact be an exact analytical expression for
the simulation error when using this resource state.

\begin{figure}[ptb]
\begin{center}
\includegraphics[
width=10cm
]{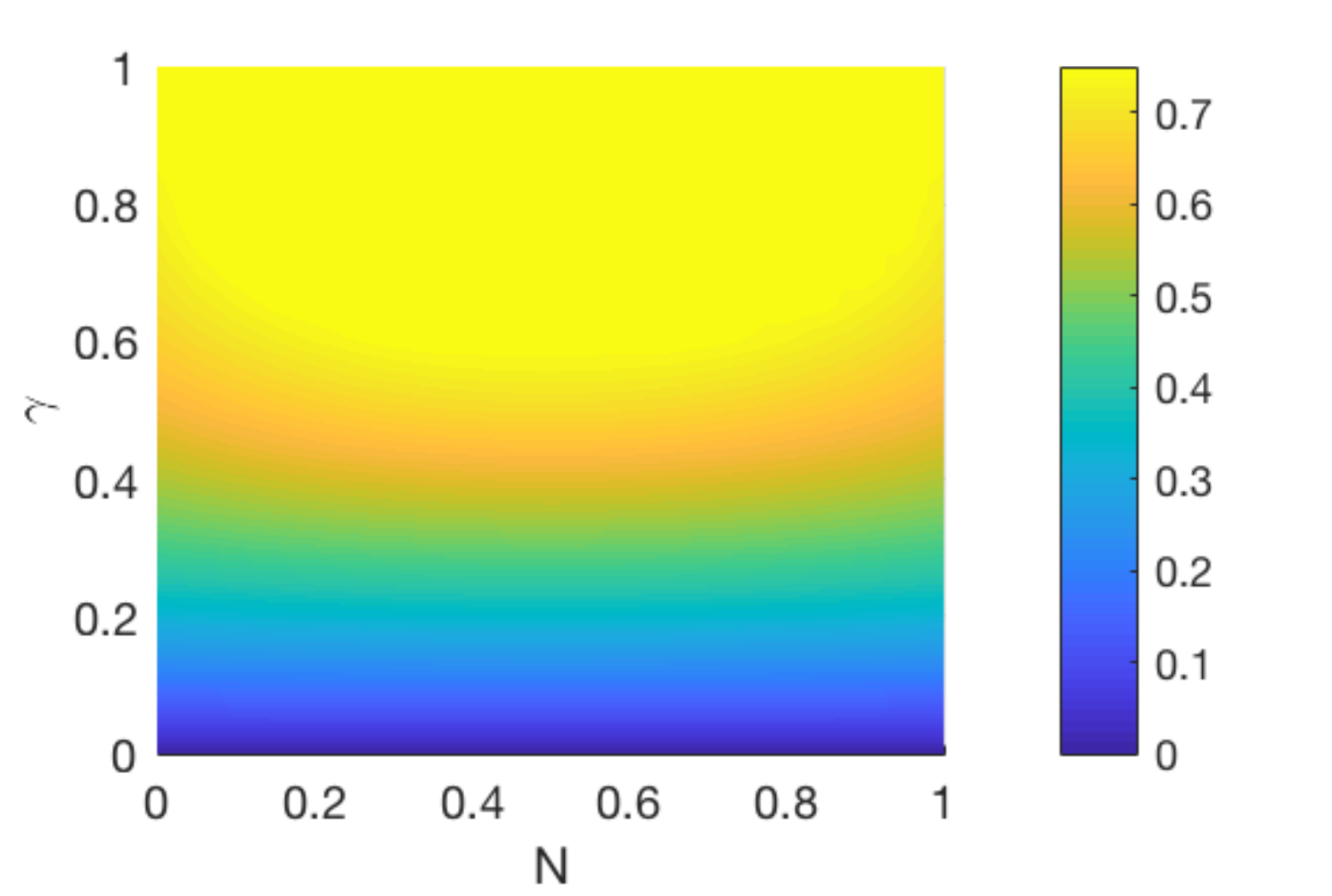}
\end{center}
\caption{Plot of the simulation error of bidirectional teleportation when
using the resource state in \eqref{eq:GADC-res-state}, for all $\gamma,N\in\left[  0,1\right]  $.}%
\label{fig:gadc-values}%
\end{figure}

Figure~\ref{fig:gadc-values} plots the
expression in \eqref{eq:upper-bnd-GADC} for the simulation error. The simulation error tends to zero as the damping parameter $\gamma$ approaches zero (so that the channel $\mathcal{A}_{\gamma, N}$ is converging to an identity channel and thus the resource state to two ebits). For fixed $\gamma$ and the noise parameter $N$ converging to 1/2, the simulation error increases.

\section{CONCLUSION}

In this paper, we provided a systematic approach
for quantifying the performance of bidirectional teleportation. We
established a benchmark for classical versus quantum bidirectional teleportation, and we have evaluated
semi-definite programming lower bounds on the simulation error for some key examples of resource states. 
Going forward from here, there are several avenues for
future work. First, we can consider other unitary channels besides the swap channel, and the line of thinking
developed here could be useful for related scenarios considered in \cite{soeda2011, wakakuwa2019}. Rather than just impersonating the swap, we can pick many other quantum channels to impersonate.
We can also consider applying the framework used here to analyze multidirectional teleportation between more than two parties. We
also wonder whether there is an LOCC simulation that
achieves a performance matching the lower bound found
here, for all parameter values for isotropic states. 

\section{ACKNOWLEDGEMENTS}

We thank Jonathan P.~Dowling for being
a catalyst for this paper. He never stopped believing in
us and our potential as researchers. His strength and
perseverance has inspired us and lives within us. May he
never be forgotten.

MMW acknowledges Moein Sarvaghad-Moghaddam
for introducing him to the topic of bidirectional teleportation. We acknowledge many insightful discussions with Justin Champagne, Sumeet Khatri, Margarite LaBorde,
Soorya Rethinasamy, and Kunal Sharma. AUS acknowledges support from the LSU Discover Research Grant,
the National Science Foundation under Grant No. OAC1852454, and the LSU Center for Computation and Technology. We also acknowledge support from the National Science Foundation under Grant No.~1907615.

\bibliographystyle{unsrt}
\bibliography{Ref.bib} 
\end{document}